 \documentclass[final,3p,times,twocolumn]{elsarticle}
 
\usepackage{amssymb}
\usepackage{ulem}
\usepackage{color}

\begin{document}

\begin{frontmatter}

\title{Quarkonia and QGP studies}

% Single author example
%
%\author{J.P Lansberg}
%\ead{lansberg@cpht.polytechnique.fr}
%\address{IPN Orsay, Paris-Sud XI, Orsay, France}

% Multiple authors
%

\author[address1,address2]{D. Blaschke}
% e-mail address
\ead{blaschke@ift.uni.wroc.pl}

\author[address1]{C. Pe\~{n}a}
% e-mail address
\ead{pena@ift.uni.wroc.pl}

%\author[address1]{D. Prorok}
% e-mail address
%\ead{prorok@ift.uni.wroc.pl}

% Address
\address[address1]{Institute for Theoretical Physics, University of Wroclaw,
50-204 Wroclaw, Poland}
\address[address2]{Bogoliubov Laboratory for Theoretical Physics, JINR,
161980 Dubna, Russia}

%\fntext[footnote1]{Footnote}

\begin{abstract}
We summarize results of recent studies of heavy quarkonia correlators and 
spectral functions at finite temperatures from lattice QCD and 
systematic T-matrix studies using QCD motivated finite-temperature potentials.
We argue that heavy quarkonia dissociation shall occur in the temperature
range $1.2 \le T_d/T_c \le 1.5$ by the interplay of both screening and 
absorption in the strongly correlated plasma medium.  
We discuss these effects on the quantum mechanical evolution of quarkonia 
states within a time-dependent harmonic oscillator model with complex 
oscillator strength and compare the results with data for 
$R_{\rm AA}/R_{\rm AA}^{\rm CNM}$ from RHIC and SPS experiments.
We speculate whether the suppression pattern of the rather precise NA60 data
from In-In collisions may be related to the recently discovered $X(3872)$ 
state. Theoretical support for this hypothesis comes from the cluster expansion
of the plasma Hamiltonian for heavy quarkonia in a strongly correlated medium. 
\end{abstract}


\end{frontmatter}

%% main text (length: 6 pages)
\section{Introduction}
\label{Introduction}
In Physics we know fields of research which are opened by ingenious 
experimental discoveries and sometimes have to wait even decades for their 
proper theoretical explanation. 
This was so in the case of superconductivity, where after Kamerlingh-Onnes' 
observation of the vanishing resistance in Mercury at liquid He temperatures 
\cite{Kamerlingh-Onnes:1911} more than four decades of theory development were 
necessary before a satisfactory explanation could be formulated 
by Bardeen, Cooper and Schrieffer \cite{Bardeen:1957kj}. 
Sometimes it is vice-versa, like in the case of Bose-Einstein condensation
\cite{Bose-Einstein:1924}. 
The verification of the theory formulated by Bose and Einstein in 1924/25 
succeeded only 70 years later \cite{Cornell-Ketterle:1995} 
since extremely subtle experimental techniques had to be developed to cool a 
sufficient number of atoms in a trap to nano-Kelvin temperatures. 

The effect of J/$\psi$ suppression (more general, heavy quarkonium suppression)
was suggested by theory \cite{Matsui:1986dk} to be a signal of quark-gluon 
plasma formation before it was actually seen in first heavy-ion collisions at
CERN SPS and thus at first glance seems to belong to the latter class of 
discoveries. 
However, after 24 years of intense experimental research and theory 
developments, one has the impression that heavy quarkonia became a field 
where theory and experiment have to work hand in hand in order to make 
progress, like at this workshop. 
A key problem to be solved is that the usage of heavy quarkonia states as a 
probe for the diagnostics of the quark-gluon plasma (QGP) state of matter 
%expected to be formed 
in ultrarelativistic heavy-ion collisions requires the knowledge of a 
baseline, e.g., from their production and evolution characteristics in 
situations when a QGP is absent.  
For a recent review see, e.g. \cite{Rapp:2008tf} and references therein. 
Current issues in the experiment-theory dialogue are summarized, e.g., in 
\cite{Brambilla:2010cs}. 
Aspects of quarkonium production at LHC are discussed, e.g., in 
\cite{Bedjidian:2003gd,Lansberg:2008zm}.  
For these proceedings we first summarize basic theory issues on the use of 
quarkonia as probes of the QGP before outlining the aspect of the
modification of quarkonium formation by a strongly correlated QGP within a 
quantum mechanical model.

The spectrum of low-lying heavy quarkonia states is perfectly decribed by 
solutions of the Schr\"odinger equation for confining potentials of the 
Cornell-type \cite{Buchmuller:1980su}. 
While at zero temperature the Cornell potential is nicely reproduced by 
lattice QCD simulations for the change in the singlet free energy of the 
system when static color charges are inserted, it has been argued that in 
a thermal system the internal energy should be used as the potential 
instead \cite{Satz:2008zc}.
The proper form of potential to be used for studying quarkonia states in the 
medium may be a superposition of both forms depending on the interplay of time 
scales for interaction and thermal relaxation in the heat bath 
\cite{Shuryak:2004tx}.
The status of this discussion is unsettled. 
In the literature (cf. \cite{Riek:2010fk}) calculations
with both potential models derived from fits to lattice data are found whereby 
the use of the free energy as a heavy-quark potential implies a weaker binding
and thus lower temperatures for the Mott transition of heavy quarkonia states,
see also \cite{Ding:2010yz}.

\begin{figure}[!htp]
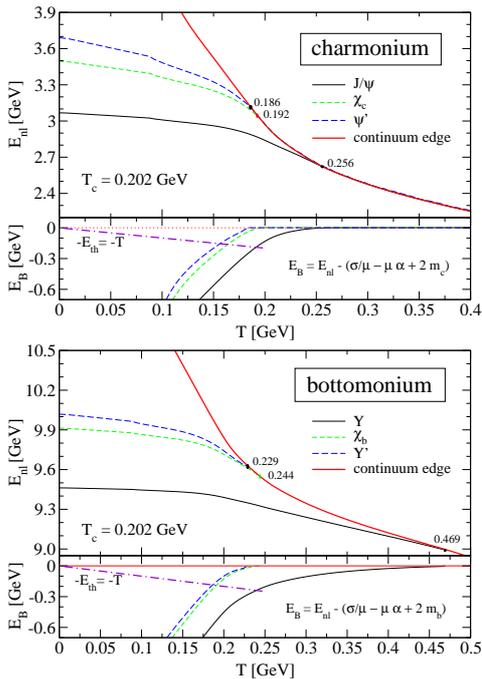

	\centering
        \includegraphics[width=0.8\columnwidth]{EB_cc.eps}
	\includegraphics[width=0.8\columnwidth]{EB_bb.eps}
	\caption{The "classical" picture: modification of binding energies of 
heavy quarkonia states in a hot plasma by static screening leads to their 
dissociation at the corresponding Mott-temperatures.  }
	\label{fig:Jpsi}
\end{figure}

In Fig.~\ref{fig:Jpsi} we illustrate the ``classical'' picture of bound state
dissociation in a hot plasma by screening of the interaction for the example 
of a screened Cornell potential applied to charmonium and bottomonium states
\cite{Jankowski:2009kr}: at state dependent (Mott) temperatures, the binding
energy vanishes and bound state merges the continuum of scattering states. 
Two comments of significance for experiments are in order: 
(i) just above the Mott temperatures, remnants of the charmonia states survive 
as resonances in the continuum to be identified by peaks in the spectral 
functions and strong correlations in the corresponding scattering phase shifts 
\cite{Blaschke:2005jg};
(ii) before reaching the Mott temperatures, the binding energies are already
sufficiently lowered so that collisions with particles from the medium may 
have sufficient thermal energy to overcome the threshold for impact 
dissociation of quarkonia \cite{Ropke:1988bx,Ropke:1988zz,Blaschke:2004dv}, 
see the lower panels of Fig.~\ref{fig:Jpsi} for 
corresponding estimates. This effect is particularly dramatic for 
$\Upsilon$ which may undergo thermal dissociation already at $T\sim 250$ MeV 
while its Mott temperature is $\sim 470$ MeV. 

Both effects tend to wash out the pattern of sequential suppression for heavy
quarkonia states expected from the ``classical'' picture of the Mott effect
\cite{Digal:2001ue}. 
The description of quarkonia states in a hot QGP medium in the vicinity of the
critical temperature should therefore treat
bound and scattering states on an equal footing. 
This is appropriately achieved within a thermodynamical T-matrix approach,
which has been developed to address the spectral properties of quarkonia 
\cite{Cabrera:2006wh} as well as open flavor meson states 
\cite{vanHees:2007me,vanHees:2008gj}. 
Such an approach allows a simultaneous description of quarkonia suppression
and heavy flavor diffusion, presently under scrutiny for RHIC experiments, see 
\cite{Riek:2010fk} and references therein.

The lowering of the quarkonia binding energy by screening has important 
consequences also for the process of quarkonium formation itself, once it 
occurs in a QGP environment. We will illustrate this aspect within a simple
model.

\section{Time-dependent harmonic oscillator model }

For our discussion of the quantum mechanical evolution of quarkonia in an 
evolving QCD plasma state, we will employ here a generalization of the 
harmonic oscillator model \cite{Matsui:1989ig} to time-dependent one with
complex oscillator strengths (THO model). 
Aspects of an optical potential for the propagation of charmonia through a 
medium have been already discussed, e.g., for cold nuclear matter in 
\cite{Kopeliovich:2003cn,Koudela:2003yd} and for a quark-gluon plasma in
\cite{Cugnon:1993ye,Cugnon:1993yf}.
The merit of such a model is its simplicity and transparence as well as 
tractability.
We consider the time-dependent Hamiltonian for heavy quarkonia in the form
\begin{equation}
H(\tau)=2m_Q+\frac{p^2}{2\mu}+\frac{\mu}{2}\omega^2(\tau) r^2(\tau)~,
\label{hamilton}
\end{equation}
where $\mu=m_Q/2$ is the reduced mass and $m_Q$ the heavy quark mass. 
The complex oscillator strength 
$\omega^2(\tau)=\omega_R^2(\tau)+i\omega_I^2(\tau)$ has an implicit time 
dependence due to the 
%dependence on the temperature of the medium and thus reflects the 
temperature evolution $T(\tau)$ of the system surrounding the evolving 
heavy quarkonium state.   
The temperature dependence of $\omega_R(T)$ resembles screening or 
strengthening of the quarkonium interaction, while a nonvanishing 
$\omega_I(T)$ signals for quarkonium absorption or dissociation processes.
The quadratic dependence of the imaginary part of the (optical) 
oscillator potential is motivated by the phenomenon of color transparency,
see also \cite{Blaschke:1992pw} and references therein. 
The conditions $\omega_I(0)=0$ and $\omega_R(0)=\omega_\psi$ apply for the
vacuum, where the spectrum of the low-lying quarkonium states is approximated 
by a suitably chosen oscillator strength $\omega=\omega_\psi={\rm const}$.

The general classical trajectories for the Hamiltonian 
(\ref{hamilton}) are linear combinations of the two solutions
%%%%%%%%%%%%%%%%%%%%%%%%%%%%%%%%%%%%%%%%%%%%%%%%%%%%%%%%%%%%%%%%%%%%%%%%%%%%%
\begin{eqnarray}
\label{e2}
r(t)=\rho(t)\exp(\pm i\phi(t))~,~~ 
%\end{eqnarray} 
%where
%\begin{equation}
\phi(t)=\int^t_{t_i}\frac{ dt'}{\rho^2(t')}~.
\end{eqnarray}
The amplitude $\rho(t)$ fulfills the Ermakov equation
\begin{eqnarray}
\label{e3}
\ddot{\rho}(t)+\omega^2(t)\ \rho(t) -\frac{1}{\rho^3(t)}=0~,
\end{eqnarray}
for which exact solutions exist  \cite{polyanin:handbook},
allowing to evaluate the time evolution operator using path integral 
methods \cite{kleinert}
\begin{eqnarray}
\label{evolution}
U(t_f;t_i)&=&\left[\frac{\mu~\rho_{f}~ \rho^{-1}_{i} \dot{\phi}_f}{2\pi
i \sin(\phi_f-\phi_i)}\right]^{1/3}\;{\rm e}^{i S_{\rm cl}}~,
\end{eqnarray}
where the classical action functional $S_{\rm cl}[\rho(t)]$ enters.
The survival probability (suppression factor) for J/$\psi$ as defined in 
\cite{Matsui:1989ig} can be generalized to the THO case and applied to the 
QGP diagnostics in collisions of heavy ions with mass number A when 
identified with the experimentally determined quantity
\begin{eqnarray}
\label{suppression}
\frac{R_{\rm AA}}{R_{\rm AA}^{\rm CNM}}&=&
\Bigg|\frac{\rho_f/\rho_i}
{\cos(\phi_f)+\left(\frac{\dot{\rho}_f}{\rho_f\dot{\phi}_f}
+i\;\frac{\omega_\psi}{\dot{\phi}_f}\right)\; \sin(\phi_f)}\Bigg|^{3}\,
\end{eqnarray}
where $R_{\rm AA}^{\rm CNM}$ accounts for the cold nuclear matter (CNM) 
effects from charmonium absorption and modification of charm production by 
shadowing/antishadowing of gluon distribution functions  in the CNM of the 
colliding nuclei. 
Both effects have been discussed in contributions to this workshop and are 
in principle accessible by analysis of pA collision experiments, see
\cite{Rapp:2008tf,Ferreiro:2009ur} and references therein.
We restrict our discussion here to ground state charmonium at rest in the QGP 
medium ($p_T=y=0$) so that the discussion of Lorentz boost effects on the 
formation process can be omitted and also a detailed discussion of feed-down
from higher charmonia states will be given elsewhere \cite{Pena:2011}. 

\begin{figure}[htb]
	\centering
	\includegraphics[width=0.8\columnwidth]{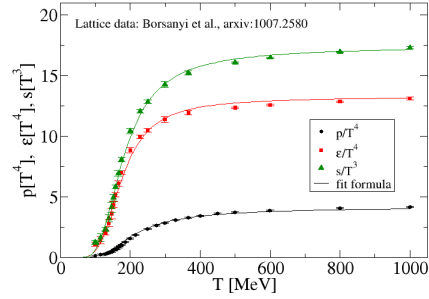}
	\caption{Recent results (data points) for the equation of state from 
lattice QCD \cite{Borsanyi:2010cj} compared to the fit formula (\ref{eos-fit})
employed here.}
	\label{fig:eos}
\end{figure}

\begin{figure}[htb]
	\centering
	\includegraphics[width=0.8\columnwidth]{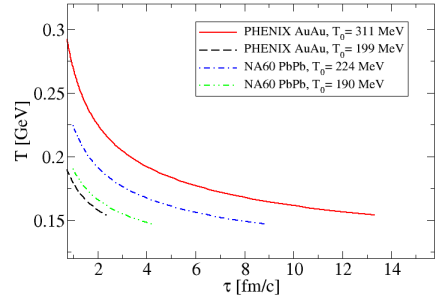}
	\caption{Temperature evolution across the QCD transition for different 
initial conditions in SPS and RHIC experiments.}
	\label{fig:hydro}
\end{figure}

In the following we show that the anomalous J/$\psi$ suppression in both,
SPS and RHIC experiments can be simultaneously described with the natural 
assumption that above the citical temperature relevant screening parametrized 
with a temperature dependent, complex oscillator strength.
The time evolution of temperature itself will be given by longitudinal (Bjorken
scaling) hydrodynamic evolution of a fireball volume $V(t)$ under entropy 
conservation
\begin{equation}
S(t)={\rm const}=s(T(t))V(t)~;~~V(t)=A_T z(t)
\end{equation}
with initial conditions determined by a Glauber model for nucleus-nucleus
collisions. 
The temperature dependence of the entropy density $s(T)$ is taken 
from recent lattice QCD simulations \cite{Borsanyi:2010cj} which are well 
parametrized by the simple ansatz
\begin{equation}
\label{eos-fit}
s(T)=9.0~T^3~\left[1+\tanh\left(\frac{T-T_0}{0.54~T} \right) \right]~,
\end{equation}
with $T_0=0.189$ GeV, see Fig.~\ref{fig:eos}. 
The resulting temperature evolution is shown in Fig.~\ref{fig:hydro} for 
initial values of temperature which correspond to initial entropy densities
given by the Bjorken formula \cite{Hatsuda}
$s_0=\frac{3.6}{A_T \tau_0}\frac{dN_{\rm ch}}{dy}\big|_{y=0}$, where the 
transverse overlap area $A_T$ is taken from \cite{Adler:2004zn}.
For the initial time $\tau_0$ of the thermodynamical fireball evolution
is assumed to depend on the center of mass energy of the collision and we 
take here $0.6~(1.0)$ fm/c for RHIC (SPS) experiments.
This completes the definition of the the THO model for applications to QGP 
studies in heavy-ion collisions which we discuss in the next section. 

\section{Anomalous suppression and the In-In case}

A key indicator for QGP formation in heavy-ion collisons is 
anomalous suppression, the deviation of experimental data for the J/$\psi$ 
production ratio (\ref{suppression}) from unity.  
This effect, first observed at CERN SPS for Pb-Pb collisions at $\sqrt{s}=17$
GeV, has been qualitatively confirmed by RHIC experiments with Au-Au 
interactions at $\sqrt{s}=200$ GeV whereby surprising new findings were 
revealed: (i) the suppression is stronger at forward and backward rapidities
rather than at midrapidity where the particle densities are highest, 
(ii) the onset of anomalous suppression and its dependence on centrality 
scales with the charged particle density at midrapidity rather than with energy
density.
While (i) is caused mainly by antishadowing and to some extent by geometry
\cite{Prorok:2009ma}, the second finding is still not understood.
A third puzzling issue raised by Carlos Louren\c{c}o in discussions at this 
meeting is (iii) the dip in the centrality dependence of the J/$\psi$ 
suppression ratio which is seen in the rather precise data of the NA60 
collaboration for In-In collisions and so far widely ignored by theorists.

We want to report rather fresh results within the THO model which might get
substantially improved in near future but the main idea of which is in the
spirit of this workshop, namely to provide a more appropriate theoretical 
basis for future experiments on heavy quarkonium production, in particular in
the LHC era. We are convinced that any deeper insights into the quantum 
mechanical evolution of the $\bar{c}c$ state in the hot QGP medium can become
very important in this context.
Here we will apply the THO model to extract the real (imaginary) part of the 
oscillator strength and thus the amount of screening (absorption) as a 
function of the temperature of the QGP medium from the experimental data.

\begin{figure}[htb]
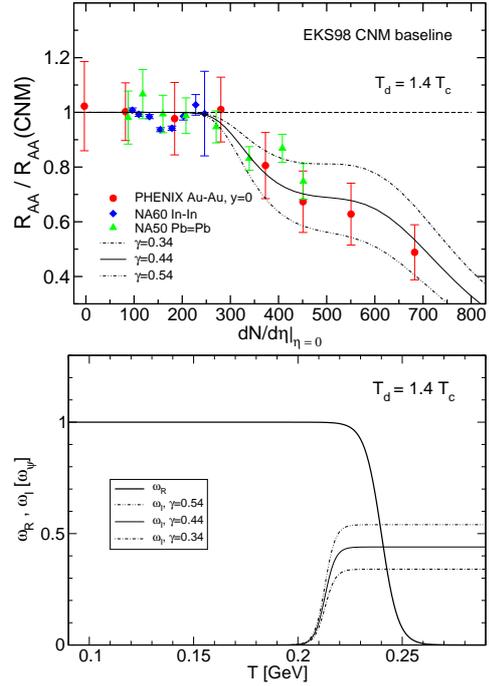

	\centering
	\includegraphics[width=0.8\columnwidth]{raa_dndeta_td14_n.eps}
	\includegraphics[width=0.8\columnwidth]{frequency_td14_n.eps}
	\caption{Anomalous J/$\psi$ suppression within the THO model compared 
to data from NA50, NA60 and PHENIX (upper panel) and the temperature dependence
of the real and imaginary parts of the HO frequency (lower panel).}
	\label{fig:Jpsi2}
\end{figure}
In Fig.~\ref{fig:Jpsi2} we present a fit which ignores the In-In dip and 
is governed by the screening of the confining interaction which monotonously 
drops to zero in the temperature range $T/T_c\sim 1.2 - 1.6$, in gross
accordance with the analysis of the charmonium spectrum from the 
temperature dependent potential on the basis of the heavy-quark free energy
from lattice QCD simulations with a Mott temperature below $1.4~T_c$.
This fit is rather insensitive to strong variations of the absorptive part of
the potential.
For recent systematic, quantum field theoretic motivation of complex 
charmonium potentials see, e.g., Refs.~\cite{Beraudo:2007ky,Laine:2007qy}
  
\begin{figure}[htb]
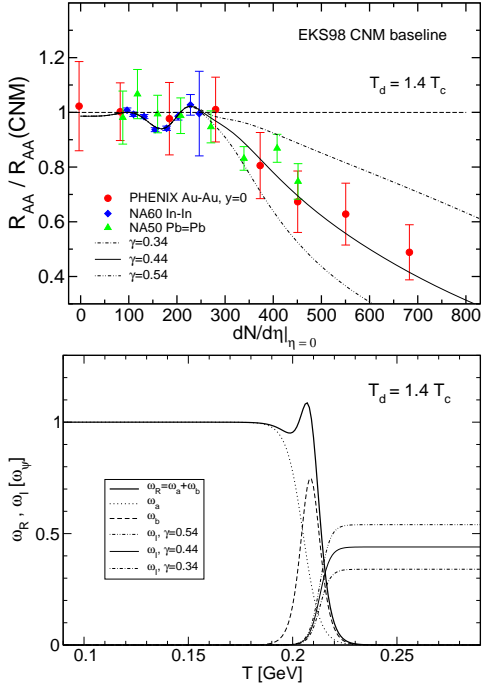

	\centering
	\includegraphics[width=0.8\columnwidth]{raa_dndeta_td14_wiggle_n.eps}
	\includegraphics[width=0.8\columnwidth]{frequency_td14_wiggle_n.eps}
	\caption{Same as Fig.~\ref{fig:Jpsi2}, but with a temperature
dependence of the HO frequency which exhibits a resonance-like structure
(lower panel), allowing for a fit of the ``dip'' in the In-In data (upper 
panel). For details, see text. 
%Anomalous J/$\psi$ suppression within the THO model compared to data from 
%NA50, NA60 and PHENIX (upper panel) and the temperature dependence
%of the real and imaginary parts of the HO frequency (lower panel).
}
	\label{fig:Jpsi3}
\end{figure}

In Fig.~\ref{fig:Jpsi3} we present a possible parametrization of the complex
THO model accounting for the dip in the In-In data 
\cite{Arnaldi:2007zz,Arnaldi:2009} while still being 
compatible with the NA50 Pb-Pb and the PHENIX Au-Au data which have 
considerably larger error bars. 
The dip reflects a nonmonotonous temperature behaviour of the confining 
potential due to a resonance-like contribution to the oscillator strength which
results in a J/$\psi$ regeneration pattern as a function of the charged 
particle multiplicity. This parametrization is more susceptible to changes in
the absorptive part in the complex oscillator strength: 
if $\omega_I(T)\stackrel{>}{\sim} \omega_\psi/2$ a too strong J/$\psi$ 
suppression would result.   

Let us present a speculation about the possible origin for a resonant-like 
strengthening of the heavy-quark potential in the vicinity of the chiral/
deconfinement transition.  
%\begin{figure}[!htp]
%	\centering
The Lippmann-Schwinger equation for $\rho$-J/$\psi$ scattering in 
the $D-\bar{D}^*$ channel can be depicted in the following way
\\[5mm]
%\centerline{\includegraphics[width=0.8\columnwidth]{T_X.png}}
\centerline{\includegraphics[width=0.8\columnwidth]{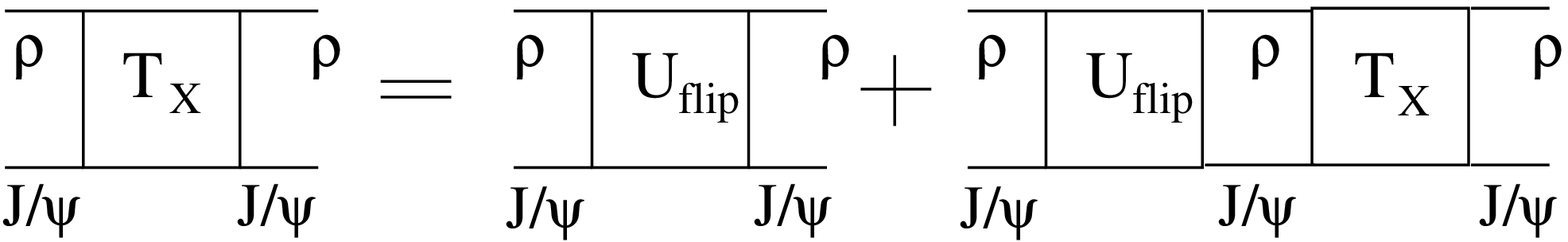}}
\\[2mm]
and its solution may produce a new state or resonance, such as the $X(3872)$
recently discovered at Belle and BaBar. 
For a recent discussion, see \cite{Brambilla:2010cs,Burns:2010qq}, and 
references therein.
The scattering kernel, 
\\[5mm]
%\centerline{\includegraphics[width=0.8\columnwidth]{U_X.png}}
\centerline{\includegraphics[width=0.8\columnwidth]{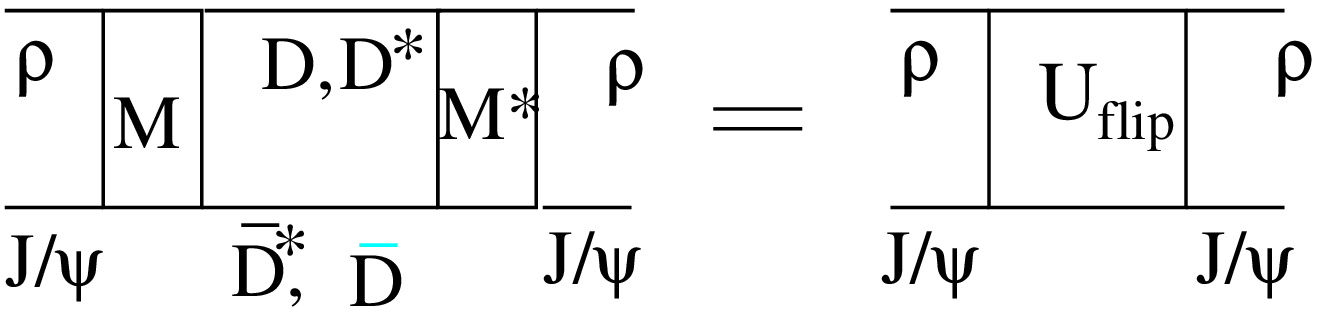}}
\\[2mm]
is given by double quark exchange.
It couples both sides of the $X(3872)$ medal: the $D-\bar{D}^*$ side which has
led to the hypothesis it may  be a molecule made of these states, and the 
$\rho$-J/$\psi$ side, the channel to which it actually predominantly decays.
Here we want to make the contact with plasma physics where a similar process 
contributes to the plasma Hamiltonian for two-particle states in a strongly 
correlated, nonideal plasma. It is represented diagrammatically as 
\\[5mm]
\centerline{
\includegraphics[width=0.4\columnwidth]{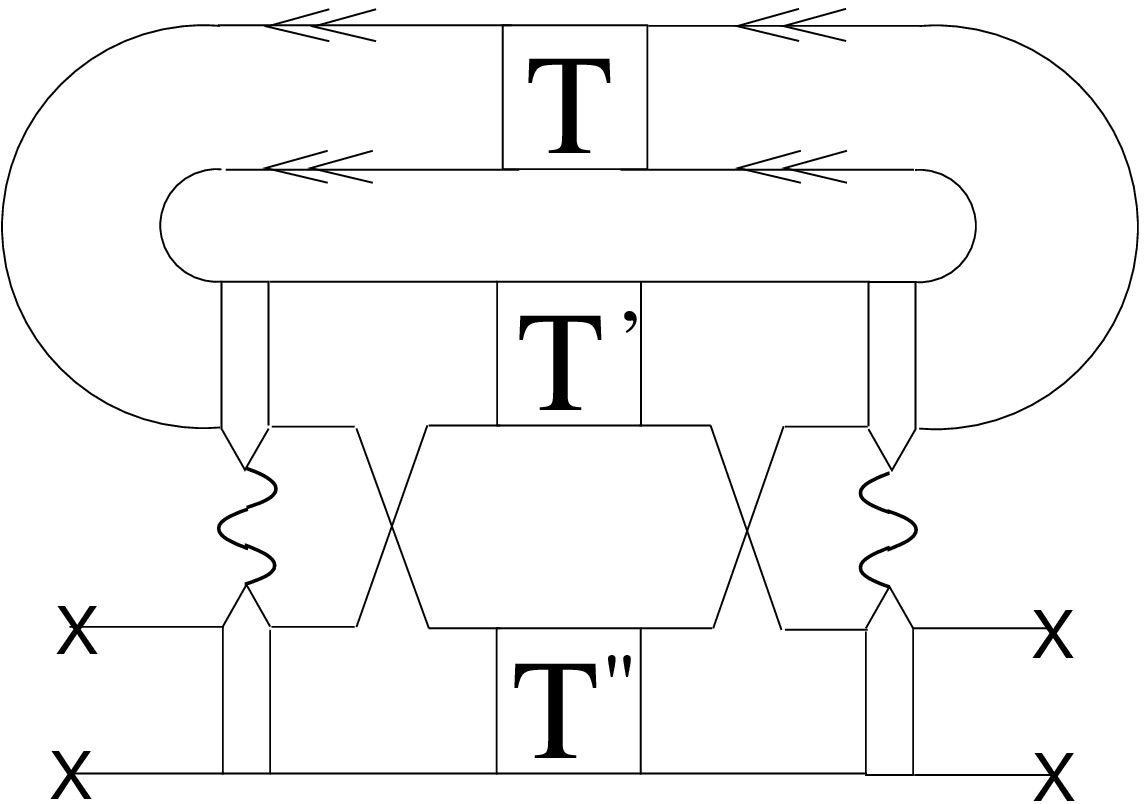}
+\dots =
\includegraphics[width=0.4\columnwidth]{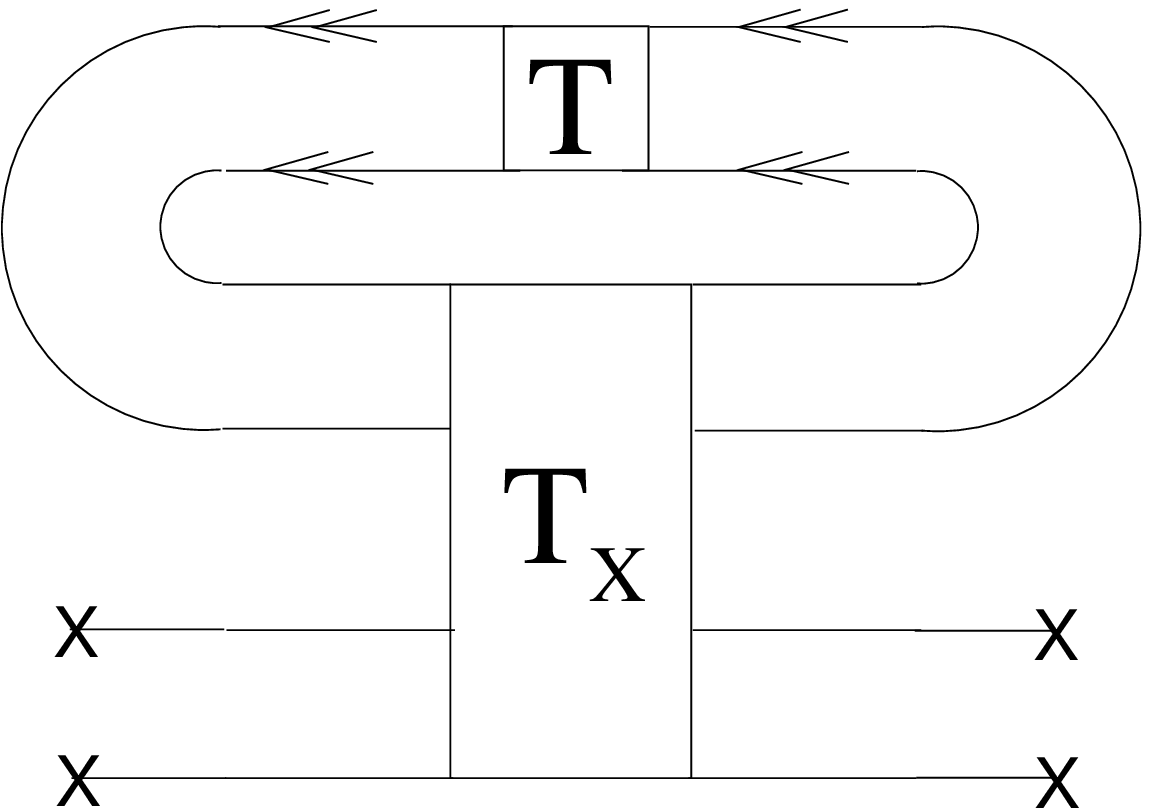}
}
\\[2mm]
and follows from a cluster expansion of the plasma medium, for details see 
Ref.~\cite{Blaschke:2009uh}. 
In the present context, we may identify the T-matrices with resonant mesonic 
states: $T=\rho$, $T'=D$ and $T''=\bar{D}*$.
The resulting contribution to the plasma Hamiltonian for quarkonia states is
proportional to the partial density of $\rho$ mesons in the medium which is 
peaked just above the QCD transition temperature, so that a contribution with 
the shape of $\omega_b(T)$ in the lower panel of Fig.~\ref{fig:Jpsi3} could be 
expected. A more detailed calculation could be based in 
Refs.~\cite{Blaschke:1992qa,Martins:1994hd} and is in progress 
\cite{Pena:2011}. 
Quark exchange processes for quarkonium suppression have first been introduced
within nonrelativistic potential models \cite{Martins:1994hd} and successfuly 
been applied to describe CERN SPS data \cite{Wong:1999zb}. 
Their reformulation within relativistic quark models has confirmed results 
for the the behaviour and magnitide of the J/$\psi$ dissociation cross section 
by $\pi$ \cite{Ivanov:2003ge} and $\rho$ meson impact \cite{Bourque:2008es}.
These models can be used to derive formfactors for the otherwise very 
powerful chiral Lagrangian approaches to charmonium dissociation, see
 \cite{Blaschke:2008mu} and references therein.
In the context of the $X(3872)$ conjecture the finding of 
Ref.~\cite{Burau:2000pn} is important that the spectral broadening of D-mesons
due to their Mott effect at the chiral transition entails a qualitative 
increase in the dissociation rate by meson impact.
Here the $\rho$ meson plays the dominant role \cite{Blaschke:2002ww},
which motivates the conjecture that the ladder-type iteration of the quark 
exchange interaction kernel $U_{\rm flip}$ may provide
sufficient strength to produce the $X(3872)$ state in that channel.

\section{Conclusion} 
Recent studies of heavy quarkonia correlators and 
spectral functions at finite temperatures in lattice QCD and 
systematic T-matrix approaches using QCD motivated finite-temperature 
potentials support that heavy quarkonia dissociation shall occur in the 
temperature range $1.2 \le T_d/T_c \le 1.5$ whereby the interplay of both 
screening and absorption processes is important.  
We have discussed these effects on the quantum mechanical evolution of 
quarkonia states within a time-dependent harmonic oscillator model with 
complex oscillator strength and compared the results with data for 
$R_{\rm AA}/R_{\rm AA}(CNM)$ from RHIC and SPS experiments.
Besides the traditional interpretation, with a threshold for the onset of 
anomalous suppression by screening and dissociation kinetics at 
$d N_{\rm ch}/d\eta \sim 300$, we suggest an alternative arising from the 
attempt to model the dip the suppression pattern of the rather precise NA60 
data from In-In collisions at $d N_{\rm ch}/d\eta \sim 150 - 250$.
We suggest that this dip indicates the true threshold for the onset of 
anomalous suppression due to the coupling of charmonium to the $\bar D_0^*D_0$ 
channel with the recently discovered $X(3872)$ state. 
Although details need to be worked out, the theoretical basis for supporting 
this hypothesis has apparently been developed in plasma physics with the 
concept of a plasma Hamiltonian for nonrelativistic bound states like heavy 
quarkonia when they are immersed in a medium dominated by strong correlations 
like bound states, to be systematically addressed within cluster expansion
techniques. 
This illustrates that in the studies of the interrelation of quarkonia with 
the QGP there are still many interesting and challenging aspects to be 
clarified within the further development and close collaboration of 
theory and experiment.   

\section*{Acknowledgments}
We would like to thank to R.~Arnaldi, E.~Scomparin and M.~Leitch for providing 
us with data shown in Figs.~\ref{fig:Jpsi2} and \ref{fig:Jpsi3}.
This work was supported in part by the Polish Ministry for Science and Higher 
Education under grants NN 202 0953 33 and NN 202 2318 37.

\end{document}